\documentclass{iopart}

\usepackage[english]{babel}
\usepackage{amssymb}
\usepackage{mathrsfs}
\usepackage[dvips]{graphicx}
\usepackage{epsfig,bm}

\newcommand{\bitem}{\begin{itemize}}
\newcommand{\fitem}{\end{itemize}}
\newcommand{\beq}{\begin{equation}}
\newcommand{\eeq}{\end{equation}}
\newcommand{\beqa}{\begin{equation} \begin{array}{rcl}}
\newcommand{\eeqa}{\end{array} \end{equation}}

\newcommand{\Dp}{\Delta p}

\begin{document}


\title[Experimental perspectives for long-range interactions]{Experimental perspectives for systems based on long-range interactions}

\author{R.~Bachelard$^1$, T.~Manos$^2$, P.~de~Buyl$^3$, F.~Staniscia$^{4,5}$, F.~S.~Cataliotti$^{2,6}$, G.~De~Ninno$^{1,4}$, D.~Fanelli$^2,7$, N.~Piovella$^8$}

\address{$^1$ University of Nova Gorica, School of applied sciences, Vipavska 11c, SI-5270 Ajdovcina, Slovenia
\\ $^2$ Dipartimento di Energetica Sergio Stecco, Universit\`a di Firenze and INFN, via S. Marta 3, 50139 Firenze, Italy
\\ $^3$ Center for Nonlinear Phenomena and Complex Systems, Universit\'e Libre de Bruxelles, CP 231, Campus Plaine, B-1050 Brussels, Belgium
\\ $^4$ Sincrotrone Trieste, S.S. 14 km 163.5, Basovizza (Ts), Italy
\\ $^5$ Dipartimento di Fisica, Universit\`a di Trieste, Italy
\\ $^6$ LENS, Universit\'a di Firenze, via N. Carrara 1, I-50019 Sesto F.no(FI), Italy
\\ $^7$ Dipartimento di Fisica, Universit\`a Degli Studi di Milano, Via Celoria 16, I-20133, Milano, Italy
\\ $^8$ Dipartimento di Fisica, Universit\`a Degli Studi di Milano, Via Celoria 16, I-20133, Milano, Italy
}
\ead{bachelard.romain@gmail.com}

\begin{abstract}
The possibility of observing phenomena peculiar to long-range interactions, and more specifically in the so-called Quasi-Stationary State (QSS) regime is investigated within the framework of two devices, namely the Free-Electron Laser (FEL) and the Collective Atomic Recoil Laser (CARL). The QSS dynamics has been mostly studied using the Hamiltonian Mean-Field (HMF) toy model, demonstrating in particular the presence of first versus second order out-of-equilibrium phase transitions from magnetized to unmagnetized regimes. Here, we give evidence of the strong connections between the HMF model and the dynamics of the two mentioned devices, and we discuss the perspectives to observe some specific QSS features experimentally. In particular, a dynamical analog of the phase transition is present in the FEL and in the CARL in its conservative regime. Regarding the dissipative CARL, a formal link is established with the HMF model. For both FEL and CARL, calculations are performed with reference to existing experimental devices, namely the FERMI@Elettra FEL under construction at Sincrotrone Trieste (Italy) and the CARL system at LENS in Florence (Italy).
\end{abstract}

\maketitle
\tableofcontents

\section{Introduction}

Long-range interactions have now been shown to be central in a wide range of scientific contexts such as astrophysics~\cite{padmanabhan90}, hydrodynamics~\cite{robert90} or nuclear physics~\cite{chomaz}. However, the possibilities of investigating the long-range features via dedicated experiments are more restricted: Non-neutral plasmas~\cite{levin08}, cold atom and wave-particle systems~\cite{coldatoms} are among the most serious candidates. The purpose of this paper is to investigate the possibility of using existing set-ups based on the wave-particle interactions to probe long-range features of the dynamics, in particular out-of-equilibrium transitions. 

As an introduction to long-range interactions, let us start from the Hamiltonian Mean-Field (HMF) model~\cite{antoni}, a paradigmatic system on which many theoretical studies focused. This one-dimensional model describes the interaction of $N$ particles on a circle through a collective field, which depends only on their phase $\theta_j$. This $N$-body dynamics is described by the following Hamiltonian:
\beq
H=\sum_{j=1}^N \left(\frac{p_j^2}{2}+\frac{\epsilon}{2N}\sum_{k=1}^N\left(1-\cos{(\theta_j-\theta_k)}\right)\right),\label{eq:hmf}
\eeq
associated to the canonical bracket in $(\theta_j,p_j)$. Here, $\epsilon=\pm 1$ corresponds either to a ferromagnetic (+) or an antiferromagnetic (-) system. In this model, the particles are collectively interacting through the so-called magnetization ${\bf M}=Me^{i\phi}=(\sum_j e^{i\theta_j})/N$, since the dynamics of a single particle is given by:
\beq
\ddot{\theta}_j+\epsilon M\sin(\theta_j-\phi)=0.
\eeq
Long-range systems can exhibit interesting equilibrium features, such as ensemble inequivalence (see e.g. \cite{dauxois00} for the antiferromagnetic two-dimensional version of the HMF model or \cite{campa09} for a recent review). However, the HMF model mainly revealed itself as a perfect playground to study {\it out-of-equilibrium} long-range features. Indeed, starting from generic non-stationary initial conditions, the system will typically have a fast transient dynamics until a nearly-stationary state, generally called Quasi-Stationary State (QSS), is reached: Not only this QSS dynamics substantially differs from the equilibrium one, but the system actually stays trapped in it for very long times~\cite{antoni}.

More specifically, several authors actually demonstrated that the lifetime of the said QSS diverges when the number of particles in interaction increases. For example, numerical works report that the time of relaxation to equilibrium for the Hamiltonian Mean-Field model scales as   $N^{1.7}$~\cite{N17}, in a regime of parameters yielding homogeneous QSS. To gain insight into the emergence of QSS, one can resort to a continuous picture, formal limit of the governing discrete Hamiltonian. A rigorous mathematical procedure leads to the Vlasov equation for the evolution of the single particle distribution function, the continuous representation of the particles density in phase space which is recovered when making the number of bodies $N$ tend to infinity. The stability of QSS in the infinite $N$ limit suggests that these latter states can be 
potentially interpreted as {\it Vlasov stationary states}, an ansatz that opens up the perspective for further analytical progress, a fact on which we shall return in the following. Operating in this context and explicitly accounting for finite size corrections beyond the idealized Vlasov picture,  the authors of~\cite{bouchet} proved rigorously that the relaxation of the $N$-body system towards its deputed equilibrium, as driven by microscopic collision effects, would occur on time scales larger than $N$, in qualitative agreement with the numerical evidences commented above. Clearly, QSS are supposedly the only regimes which are made experimentally accessible, in all physical situations where a large  number of microscopic constituents evolve in mutual interaction. The experimental time of observation is in fact limited, and not sufficient to allow for equilibration. In this perspective, to unravel the puzzle of QSS and so build a comprehensive dynamical picture for 
their existence and evolution, represents a major challenge, with undoubtedly many practical implications. 

An important step forward explaining the presence of the QSS was eventually attained thanks to the theory of violent relaxation of Lynden-Bell (LB) \cite{LB67}. This is
a statistical theory which embeds self-consistently knowledge of the governing Vlasov dynamics. The approach is justified from first principles and 
allows to resolve the intermediate regime of the discrete $N$-body evolution, when the system is presumably assimilable to a continuum Vlasov model,
before finite size corrections come eventually into play. The theory is based on the maximization of the following entropic functional of the distribution function (DF) $\bar{f}$:
\beq
s[\bar{f}]=-\int dp d\theta \left[\frac{\bar{f}}{f_0}\ln{\frac{\bar{f}}{f_0}}+\left(1-\frac{\bar{f}}{f_0}\right)\ln{\left(1-\frac{\bar{f}}{f_0}\right)}\right],
\label{entropy}
\eeq
where $f_0$ describes the initial state of the system, whereas $\bar{f}$ stands for a coarse-grained distribution function of the final state, that one wishes to recover via a predictive approach. The above formulation holds for a two-step initial distribution function (water bag): $f$ at time $0$ is equal to either zero or $f_0$. Whereas the exact evolution according to the Vlasov equation imposes that the DF is only allowed to take $0$ and $f_0$ values at all times, the coarse-grained point-of-view implies a continuous DF $\bar f$ that is expected to be valid if one averages over small patches of phase space. As a side comment we notice that the functional (\ref{entropy}) can be readily generalized to account for a continuous collection of different density levels, beyond the water-bag hypothesis. 

The maximization of $s$ is performed under the macroscopic constraints of normalization, energy and momentum which are conserved by the dynamics.
An underlying hypothesis to the theory is that the system explores in an ergodic-like fashion all states allowed by the constraints. The dynamical evolution of the Vlasov equation departs from that of a system sampling the equilibrium microcanonical ensemble, giving rise to different predictions which reflect the out-of-equilibrium nature of the problem. The application of the above predictive strategy to the study of the QSS dynamics of respectively the HMF model~\cite{antoniazzi07}, free-electron lasers~\cite{barre04} and gravitational systems~\cite{yamaguchi08} has confirmed its adequacy. 

The LB approach also brought some new insights into the HMF phenomenology. For example, the abrupt change in the QSS magnetization when smoothly tuning the initial state of the system was interpreted as an {\it out-of-equilibrium phase transition}, which not only depends on the energy of the system - as it happens at equilibrium - but also on the precise way the system is prepared. More specifically, the initial magnetization of the system was shown to have an important role, and phase transitions of both first and second order could be observed depending on the value of this latter parameter. This memory effect - the system keeps track of the detail of its initial state for very long times - makes the QSS dynamics significantly richer than the equilibrium one. For example, as regards the HMF model, an out-of-equilibrium tricritical point was identified, which does not exist at equilibrium.

The purpose of this paper is to determine whether some of the QSS features predicted for the HMF toy-model can be observed in experiments run for a dedicated class of devices. Motivated by this working hypothesis,  we shall turn to considering the wide field of wave-particle interaction and focus in particular onto two different experiments, namely the Free-Electron Laser (FEL) and the Collective Atomic Recoil Laser (CARL). In both cases the dynamics reflects the  long-range nature of the interaction, along the lines depicted above with reference to the simplified HMF setting. Operating in this framework, we will show that some features of the QSS dynamics, as those previously outlined, may be observed in direct experiments. Moreover, such properties though peculiar to the considered wave-particle dynamics, bear some reminiscent traits of the HMF model, to which both FEL and CARL are intimately connected. Eventually, the associated experimental set-ups are briefly detailed, based on existing machines and current technology.

Section \ref{sec:FEL} is devoted to FELs. The aim of such devices is to produce high-power short-wavelength light pulses by exploiting the radiation emitted by ultra-relativistic electrons when passing through the static and periodic magnetic field generated by an undulator. Starting from generic initial conditions, the wave power grows to a maximum, and then starts oscillating, keeping a lively exchange of energy with the particles, over times diverging with the number of particles, a characteristics of the QSS. As for the case of HMF, the QSS of a FEL depends not only on the energy of the system, but also on the details of its initial state. Thus, after presenting the FERMI@Elettra FEL, we discuss how to manipulate the electron beam  to produce the sought different initial states. Finally, the dynamical transition present in the system is described.

Section \ref{sec:CARL} is dedicated to discussing the Collective Atomic Recoil Laser (CARL), an experiment where a probe wave is amplified thanks to a grating of cold atoms (back)scattering photons of an incident pump laser beam. As for the FEL, its dynamics is dominated by long-range effects in the one-dimensional limit, an approximation which holds for the CARL experiment based at the European Laboratory for Non-linear Spectroscopy (LENS). We then focus on the conservative regime, when the dynamics formally reduces to that of the FEL: The possibilities to observe for the CARL the QSS phenomenology as depicted for the FEL is investigated. On the other hand, when the wave amplification takes place in a cavity, a damping has to be accomodated for: A formal link between this operational regime of CARL and the HMF dynamics is drawn, as well as the experimental perspectives to detect the associated out-of-equilibrium transitions.

Finally, in Section \ref{sec:ccl}, we discuss the measurements that could be performed for both CARL and FEL in order to unravel the imprint of QSS that indirectly materializes in the existence of distinct out-of-equilibrium regimes.

\section{The Free-Electron Laser as a long-range interacting system\label{sec:FEL}}

\begin{figure}[!ht]
\centerline{
$\begin{array}{cc}
\epsfig{figure=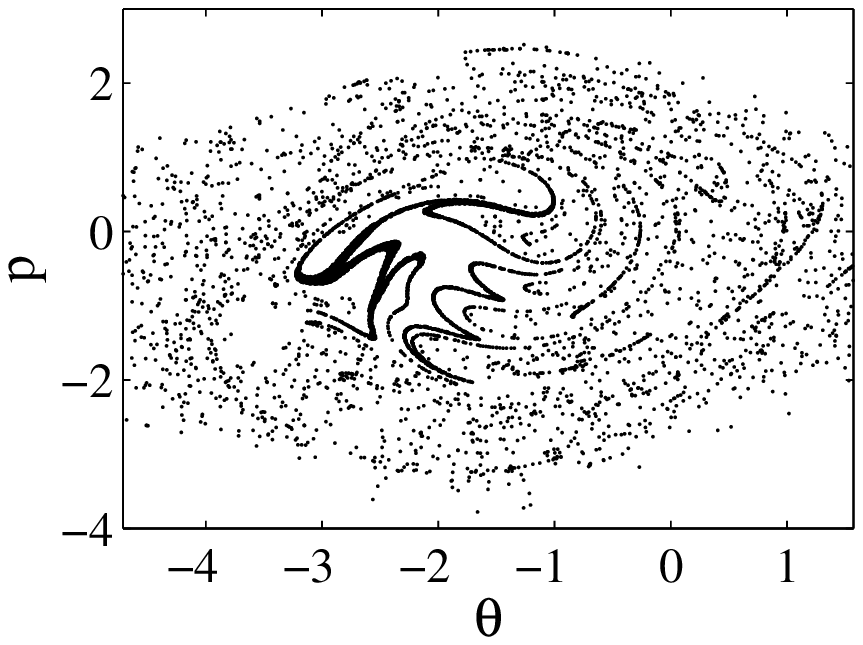,width=6cm}&\epsfig{figure=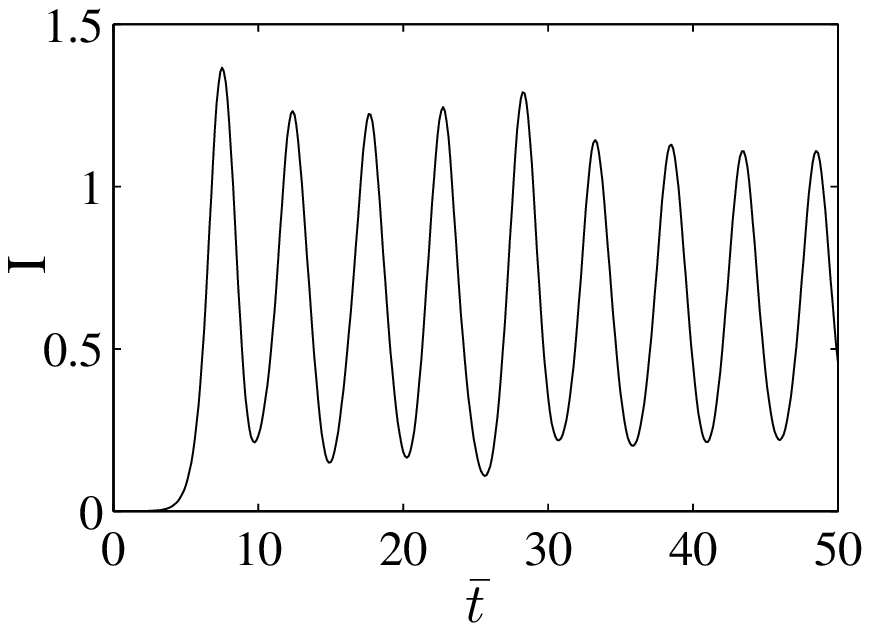,width=6cm}
\end{array}$}
\caption{Left: Electron phase-space in the QSS regime (at $\bar{t}=50$). Right: Normalized laser intensity versus normalized time $\bar{t}$. Simulations performed with $N=8000$ particles, starting with a waterbag with $b_0=0$, $\Delta p=0.05$, and a negligible intensity $I_0=10^{-6}$.\label{fig:FELdynamics}}
\end{figure}

FELs are powerful light sources able to deliver coherent pulses of photons over a large and tunable wavelength range. To that aim, ultra-relativistic electrons of energy $\gamma$ are injected into the periodic magnetic field (of period $\lambda_w$ and deflection parameter $K$) produced by an undulator, where they start to wiggle and emit synchrotron radiation around the following wavelength:
\beq
\lambda=\frac{\lambda_w}{2\gamma^2}\left(1+K^2\right).\label{eq:reso}
\eeq
The light produced by the electrons traps the electrons themselves, resulting in a periodic modulation of the electrons' density (see Fig.\ref{fig:FELdynamics}) called bunching: This bunching is the source of the coherent emission. Eventually, under a resonant condition between the electrons and the wave, the strong interplay between coherent emission and particle trapping inside the wave potential leads to the nonlinear growth of the wave (see Fig.\ref{fig:FELdynamics}) and to the emission of a powerful light pulse.
Due to the high energy of the electrons (of order 1 GeV typically), the system can be in first approximation considered as one-dimensional, since the angle of the cone of light radiated goes as the inverse of the electrons energy. As for the radiation, it can generally be described by a mean-field wave, leading to the set of equations~\cite{colsonbonifacio}:
\beqa
\frac{d\theta_j}{d\bar{t}}&=&p_j,
\\ \frac{dp_j}{d\bar{t}}&=&-\left(A e^{i\theta_j}+A^* e^{-i\theta_j}\right),
\\ \frac{dA}{d\bar{t}}&=& \frac{1}{N}\sum_j e^{-i\theta_j}+i \delta A.\label{eq:dynFEL}
\eeqa
where $\theta_j$ is the phase of electron $j$ with respect to the ponderomotive potential, $p_j$ its normalized energy, whereas $A$ stands for the complex amplitude of the synchrotron radiation. The normalized variables are defined as $\theta_j=(k+k_w)z_j-\omega t-\delta \bar{t}$, with $z_j$ the position of particle $j$ along the propagation axis, $p_j=(\gamma_j-\gamma_0)/\rho\gamma_0$, $\gamma_0$ the average electron energy, $k$ and $\omega$ the radiation wavenumber and frequency, $\delta=(\gamma_0-\gamma_R)/\rho\gamma_0$ the detuning parameter and $\gamma_R$ the resonant energy defined by Eq.(\ref{eq:reso}). $\rho=(I/I_A)^{1/3}(\lambda_w a_w/2\pi \sigma)^{2/3}/2\gamma_0$ is the so-called Pierce parameter, $a_w=\sqrt{2}K$, $I=n_e 2\pi \sigma^2 ec$ the electron current, $n_e$ the electron density and $I_A=17kA$ the Alfven current. $A$ corresponds to the normalized electric field of the wave, according to $A=E\sqrt{\epsilon_0/(mc^2\gamma_0 n_e\epsilon_0\rho)}$, while the rescaled time $\bar{t}$ is given by $\bar{t}=2k_w\rho z$, with $z$ the position along the propagation axis.

\begin{figure}[!ht]
\centerline{
$\begin{array}{c}
\epsfig{figure=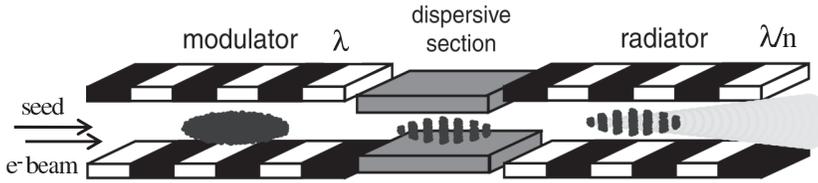,width=12cm}
\end{array}$}
\caption{Schematic layout of the HGHG scheme: The electron beam is synchronized with the seed laser, which creates an energy modulation in the former inside the modulator. In the dispersive section, the energy modulation is converted into a spatial one, called micro-bunching. In the radiator, the micro-bunched electron beam emits coherently.\label{fig:hghg}}
\end{figure}

Following the HMF approach, we focus on the waterbag initial conditions, since they are a good description of the electron bunch as a first approximation~\cite{barre04}: the initial wave is initially of zero amplitude, while the particles are bunched into an homogeneous rectangle in the $(\theta,p)$ phase-space, i.e. spread between $-\Delta \theta$ and $+\Delta \theta$ in phases, and between $-\Delta p$ and $+\Delta p$ in momenta. Experimentally, shaping the initial electron bunch is part of the High Gain Harmonic Generation (HGHG) scheme~\cite{yu91}, where use is made of two distinct stages of interaction (see Fig.\ref{fig:hghg}): In the first undulator sections, called  the ``modulator'', electrons interact with an external coherent light source, e.g., a high-power laser (called the ``seed''). Such an interaction induces the electron bunching at the seed wavelength, $\lambda_{seed}$, and at the harmonics of the latter.  In a second undulator section, tuned at one of the seed harmonics wavelength and called the ``radiator'', electrons emit coherently.

 In general, the wave grows, first quadratically, then exponentially, until it reaches a maximum and starts oscillating around an average value $\bar{I}$. As for the electrons, they bunch together, thus allowing the coherent emission; the transfer of energy to the wave leads to a decrease in the electrons energy, spoiling the resonant condition (\ref{eq:reso}). The amplification process stops when the particles are not any more in resonance with the wave (saturation). Note however that when the energy spread $\Delta p$ is too large (typically $\Delta p\geq 1.5$), the interplay between the wave and the electrons will not even trigger, and the wave amplification will not happen.

The FERMI@ELETTRA is a new FEL, presently under construction at the Sincrotrone Trieste laboratory. It aims at producing GW optical pulses in the $10-100 nm$ range, thanks to the HGHG process: In this scheme, when considering a jump from $\lambda_{seed}$ in the modulator to its $n$th harmonic ($\lambda_{rad}=\lambda_{mod}/n$, with $n$ an integer) in the radiator,the bunching at the entrance of the radiator, is given by~\cite{yu91}
\beq
|b_n|=|<e^{in\theta}>|=e^{-\frac{1}{2}n^2\sigma_\gamma^2 d^2}J_n(nd\Delta\gamma),
\eeq
where $<>$ is the average over the particles, $J_n$ the $n$-th Bessel function of the first kind, $\sigma_\gamma$ the initial energy spread of the electron beam, and $\Delta\gamma$ the ``coherent'' energy spread generated by the modulation (see~\cite{curbis07} for details, and Tab.\ref{tab:fermiprms} for the FERMI parameters). Here, $d$ is the strength of the dispersive section, whose role is to convert the energy modulation into a spatial one.

From now on, let us consider a HGHG configuration where the seed wave is $\lambda_{mod}=200nm$, associated to an harmonic jump of $n=2$ ($\lambda_{rad}=100nm$), with electron injected at $\gamma_0=1760$ into a $z=18.4m$-long radiator (see Tab.\ref{tab:fermiprms} for the FERMI FEL parameters). Then, the maximum bunching is reached when the Bessel function $J_2$ is maximized, that is for $nd\Delta\gamma\approx 3.05$, and setting the $n^2\sigma_\gamma^2 d^2$ term close to zero. Thus, a dispersive section strength of $d=0.63/(n\sigma_\gamma)$ leads to a decrease of $20\%$ in $|b_n|$ for the exponential term, and corresponds to a $\Delta \gamma\approx 3.5\sigma_\gamma$. The initial and coherent energy spreads accumulating as $\sigma_{\gamma,tot}=\sqrt{\sigma_\gamma^2+(\Delta\gamma)^2/2}$, we get $\sigma_{\gamma,tot}\approx 0.12$, whereas the bunching factor created is $|b_n|\approx 0.4$.
\begin{table}[!ht]
\caption{\label{tab:fermiprms}Main parameters of FERMI modulator and radiator sections.}
\begin{center}
\begin{tabular}{@{}cccccccc}
\br
Section & $L_w$ & K & $\gamma_0$ & $\rho$ & $\lambda$ & $\sigma_\gamma$ \\
\mr
Modulator & 3m & 1-5 & 1760-2940 & $\sim$ 3.10$^{-3}$ & 800-100nm & 0.035 \\
Radiator & 13.8-18.4m & 1-5 & 1760-2940 & $\sim$ 3.10$^{-3}$ & 100-20 nm & 0.035-0.5 \\
\br
\end{tabular}
\end{center}
\end{table}

As for the detuning, it is induced by shifting the resonant energy in the radiator from the average electron energy $\gamma_0$, according to the relation:
\beq
\delta=\frac{\gamma_{0}-\gamma_{R}}{\rho\gamma_{0}}.
\eeq

The $\bar{I}>0$ saturated regime of the FEL was shown to be accurately described by the LB approach~\cite{barre04}: The intensity and bunching reached by the laser are in good agreement with those determined by the maximization of entropy principle. It seems however not to apply to the non-lasing regime, when the resonance between the wave and the electrons is not satisfied anymore. A recent work~\cite{debuyl09} reported that regarding the LB principle, two solutions of the maximization problem exist: the one associated to a positive laser intensity is always entropically favored, but the system dynamics can actually be trapped in the vicinity of a zero-intensity solution, where the electrons stay unbunched.

\begin{figure}[!ht]
\centerline{
$\begin{array}{cc}
\epsfig{figure=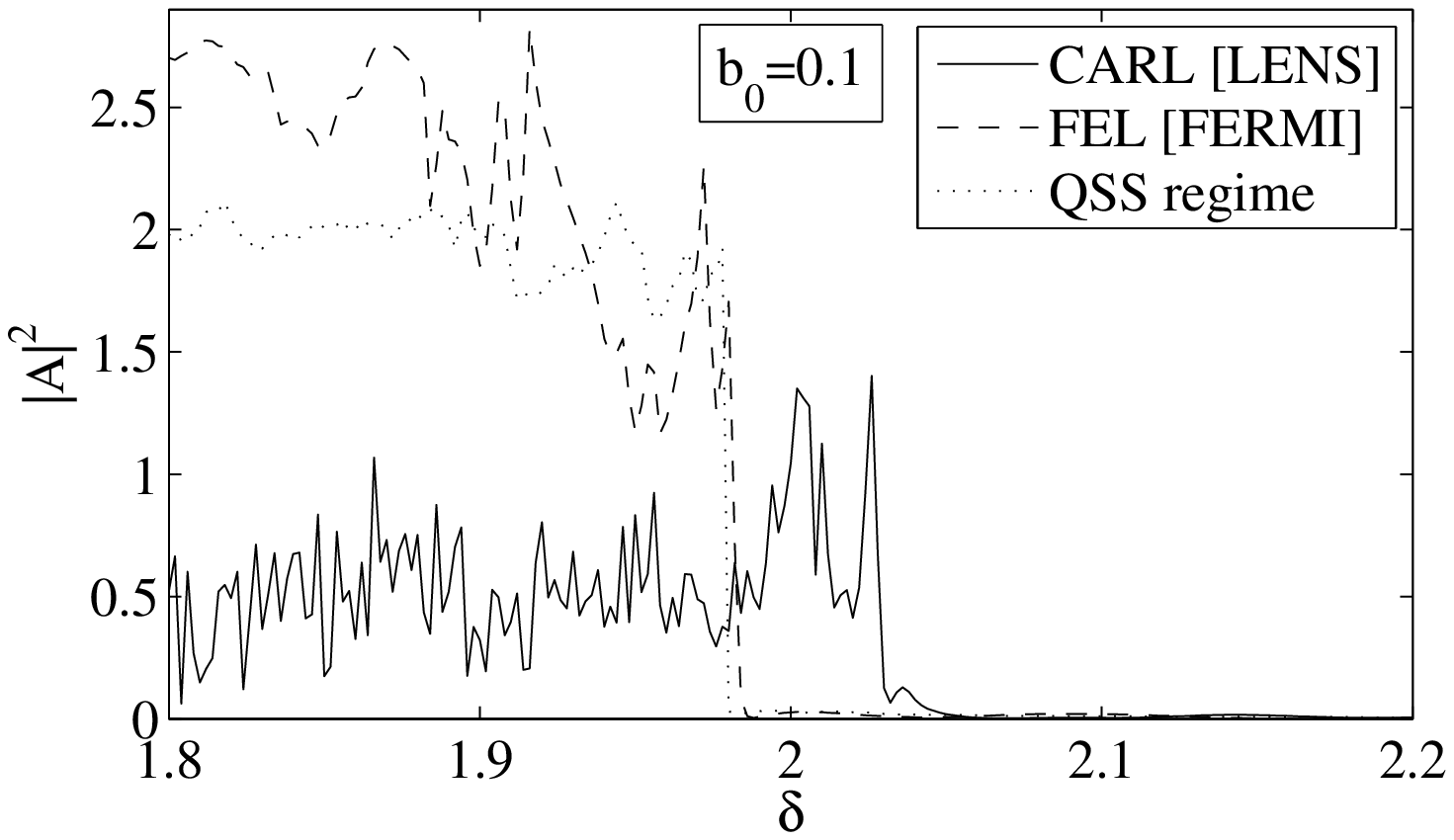,width=8cm} & \epsfig{figure=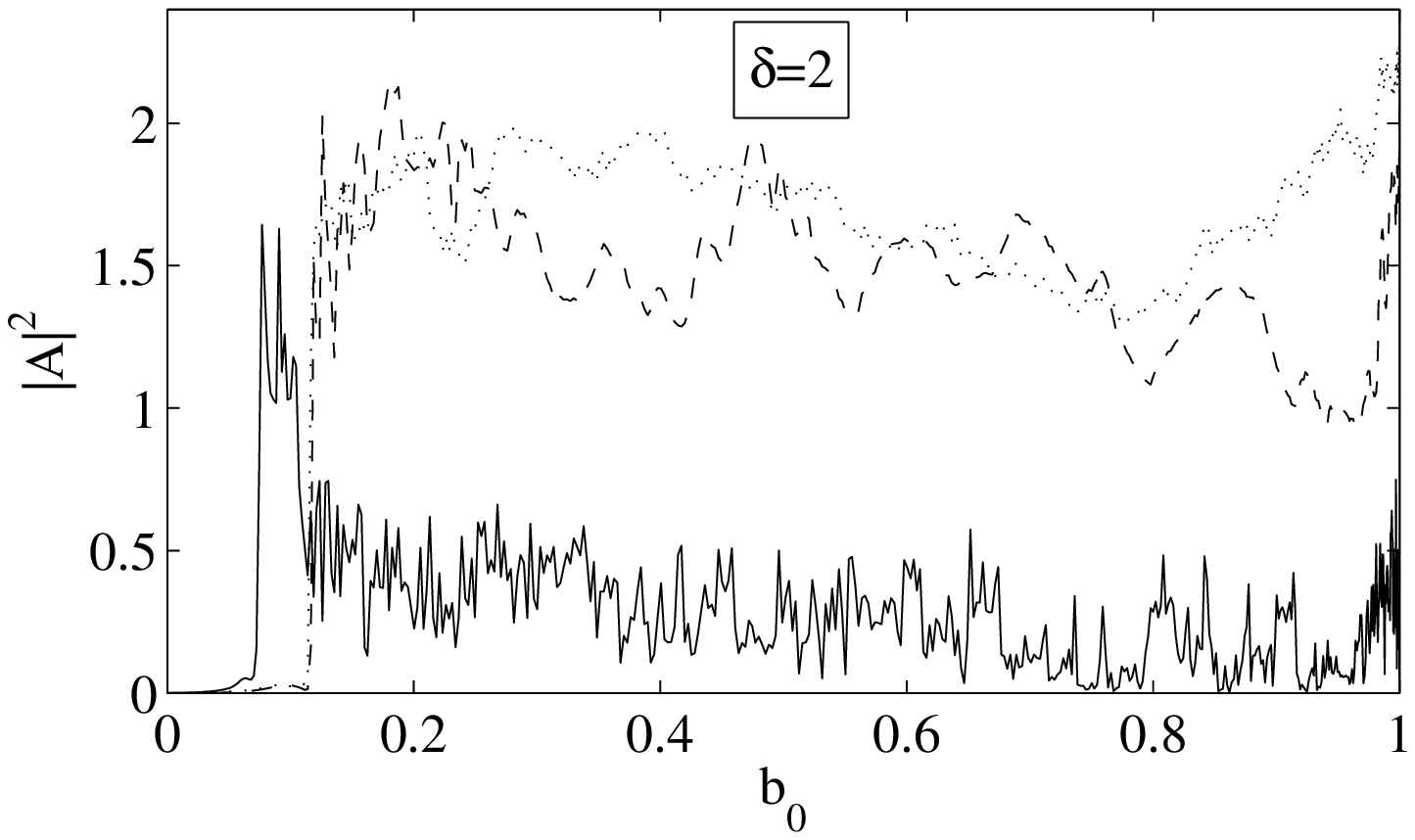,width=8cm} 
\\ \epsfig{figure=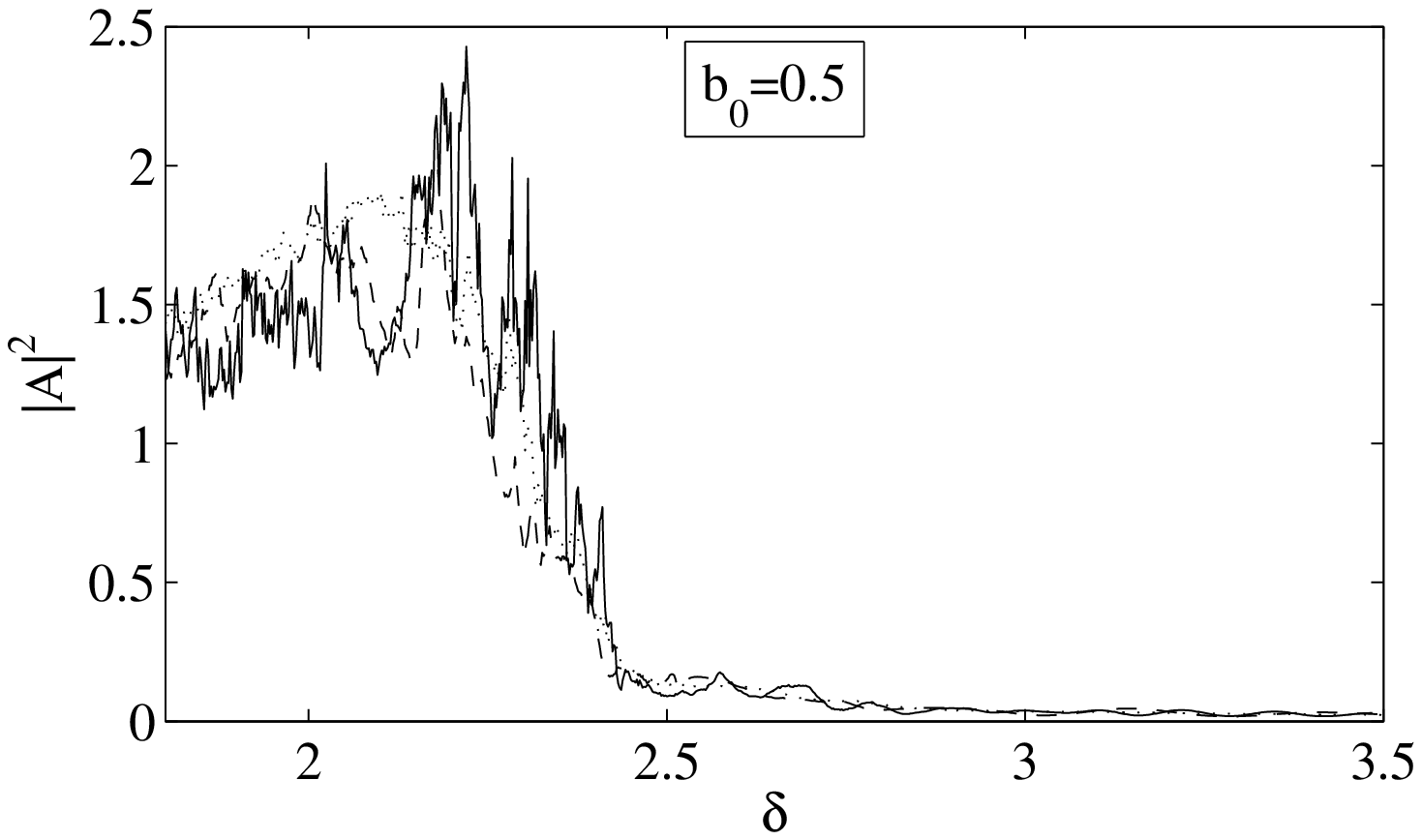,width=8cm} & \epsfig{figure=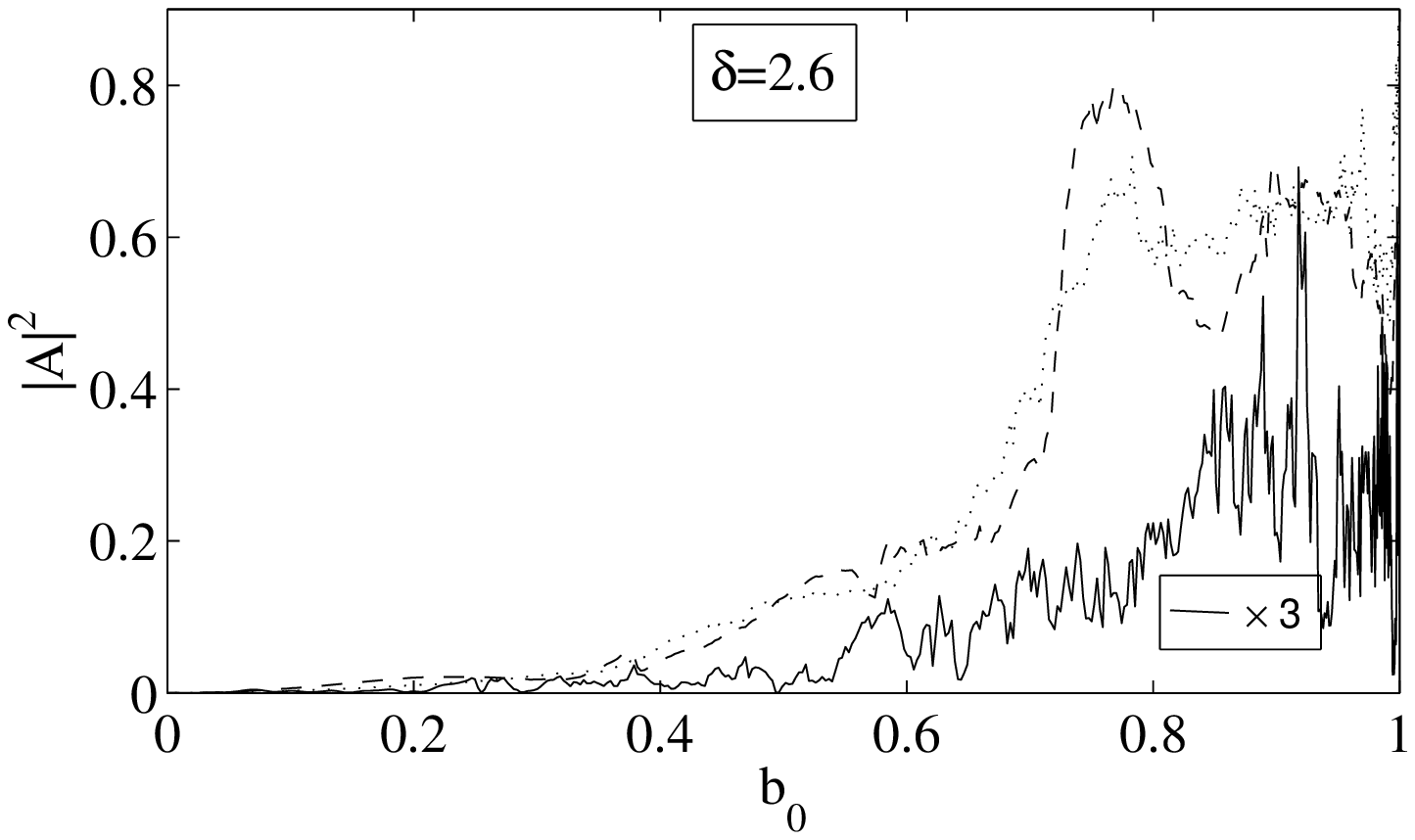,width=8cm}
\end{array}$}
\caption{Intensity radiated versus control parameter ($\delta$ on the left [at fixed $b_0$], and $b_0$ on the right [at fixed $\delta$]). The intensity for the FEL (dashed lines) are simulated using the FERMI parameters ($\gamma_0=1760$, $\lambda=100nm$, $\rho=4.35.10^{-3}$, $z=18.4m$, $\sigma_{\gamma,tot}=0.12$); for the CARL (plain lines), the LENS parameters have been used ($\rho_C=1000$, $t=1\mu s$ and energy spread $\sigma_p=20\hbar k$). The saturated intensity $\bar{I}$ of the QSS regime (dotted lines) is calculated as the average of the intensity between $t=50$ and $100$. Simulations performed with $N=10000$, and no initial wave. In the last picture, the CARL curve is represented three times larger.\label{fig:slicesFEL}
}
\end{figure}

This {\it dynamical trapping} exhibits striking similarities with the phase transition encountered in HMF. Indeed, when monitoring control parameters such as the initial bunching $b_0$ or the energy spread $\Dp$, a transition from a $\bar{I}>0$ regime to a $\bar{I}\approx 0$ one occurs, which can be sharp or smooth (see~\cite{debuyl09}). Let us focus on the parameters $b_0$ and detuning $\delta$: Fig.\ref{fig:slicesFEL} shows how the $\bar{I}>0$ regime may abruptly end for $b_0=0.1$ or values of the detuning $\delta=2$ (top panels), whereas the transition to low-$\bar{I}$ regimes is smooth for $b_0=0.5$ or $\delta=2.6$ (bottom panels). It is worth noting that working at finite undulator length does not modify substantially the transition characteristics (dashed lines of Fig.\ref{fig:slicesFEL}). 

\begin{figure}[!ht]
\centerline{
$\begin{array}{ccc}
\epsfig{figure=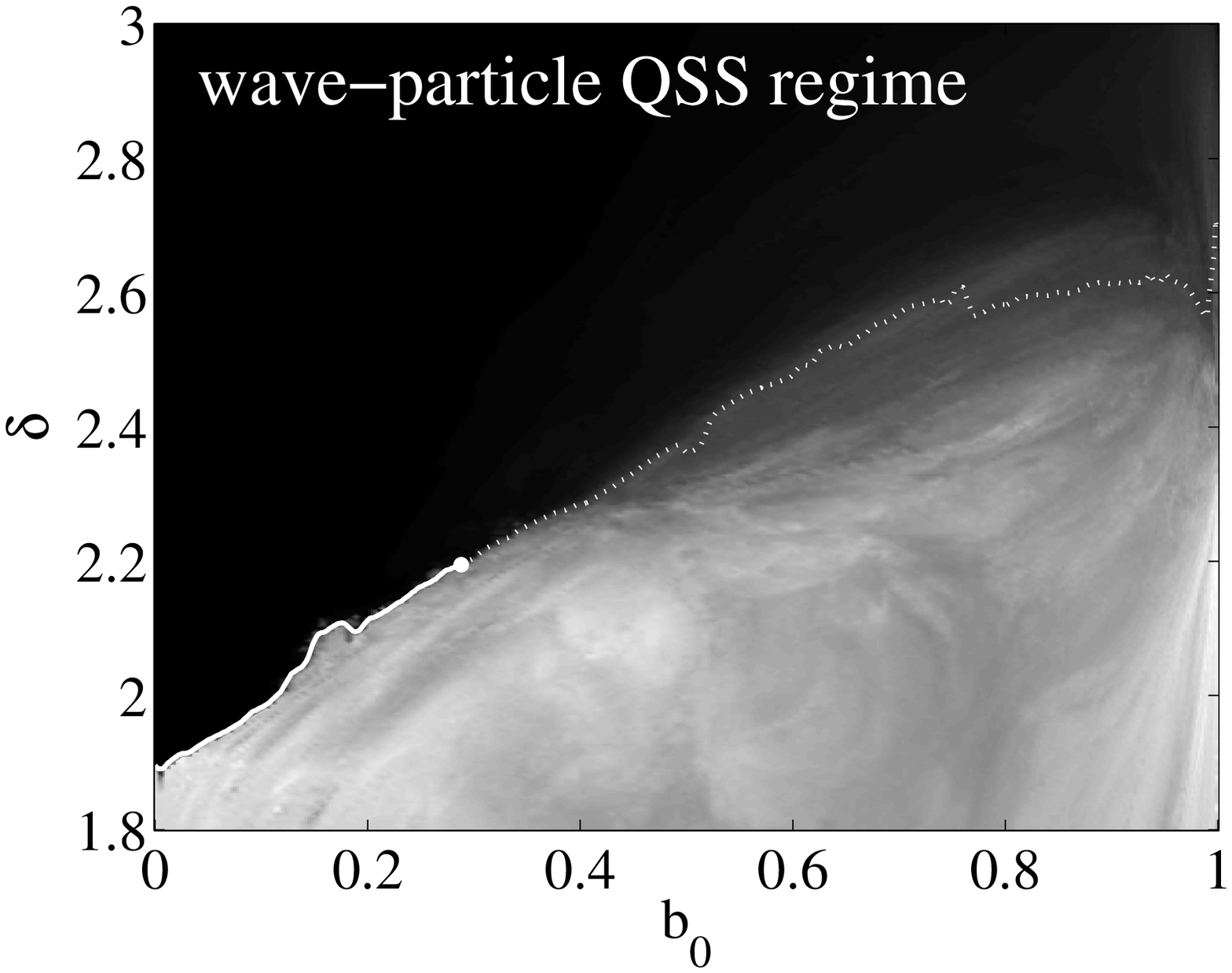,width=5cm}&\epsfig{figure=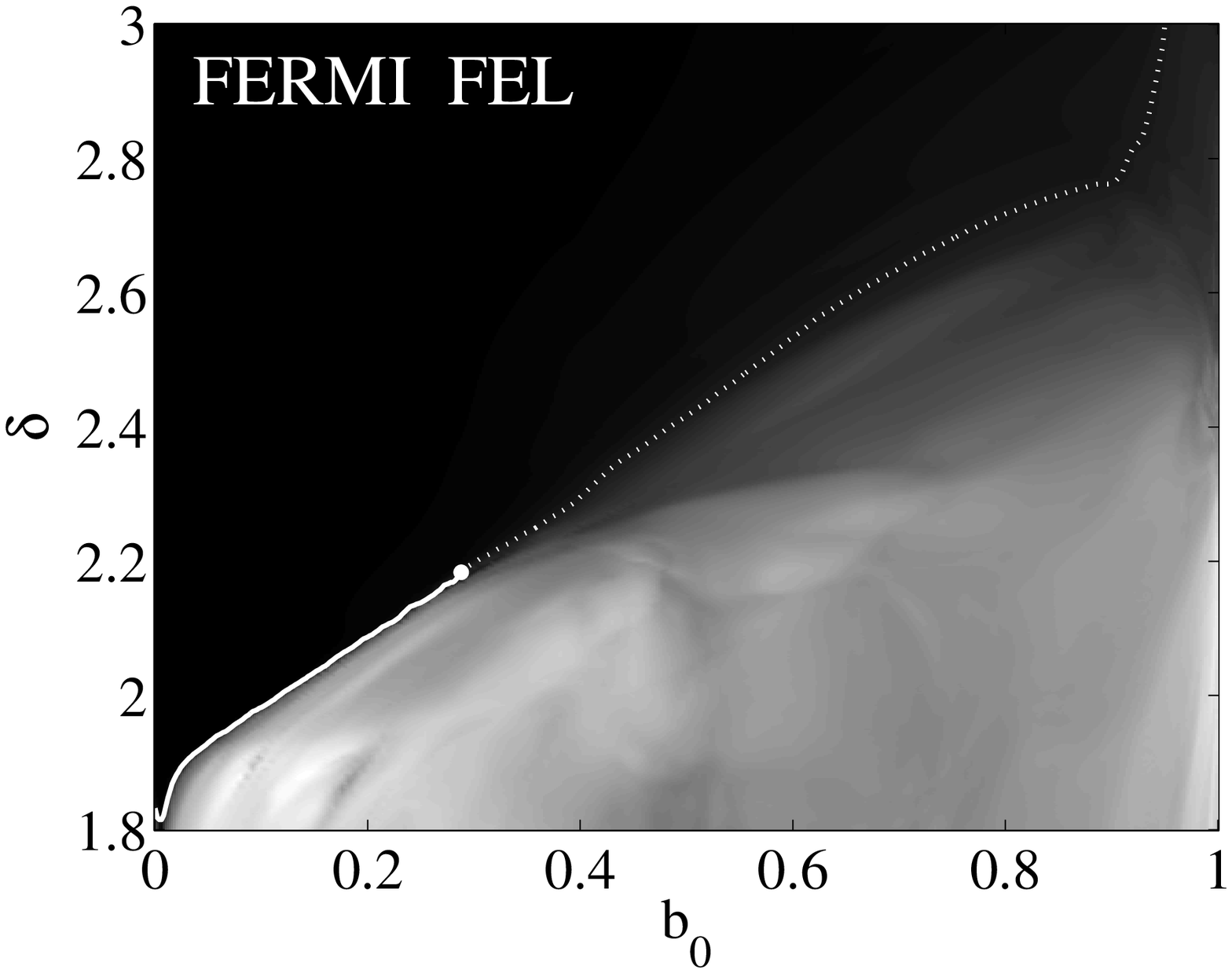,width=5cm}&\epsfig{figure=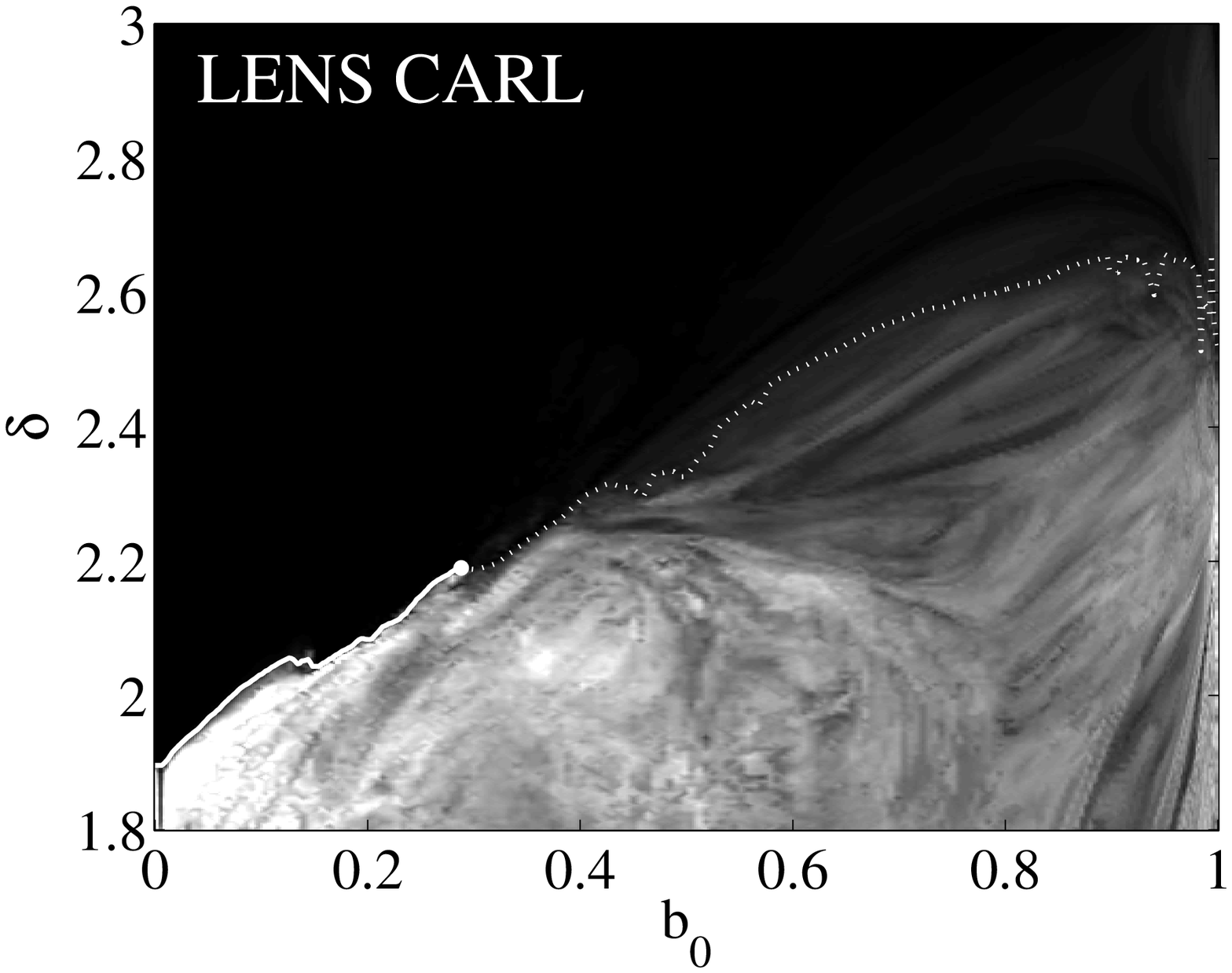,width=5cm}
\end{array}$}
\caption{Intensity as a function of $(b_0,\delta)$ in the QSS regime (left), and at the exit of the FERMI@Elettra FEL (center) and CARL of LENS (right). Bright areas correspond to high intensities, dark ones to low-intensity regimes. The white lines represent the transition between the $\bar{I}>0$ regime and the $\bar{I}\approx 0$ one, with a transition either sharp (plain line) or smooth (dashed line). The white dot stands for the tricritical point $(b_c,\delta_c)$, when the transition in the intensity goes from sharp to smooth. Simulations performed in the same conditions as Fig.\ref{fig:slicesFEL}.\label{fig:PDFEL}}
\end{figure}

The regions of parameters where each transition will occur is summarized in Fig.\ref{fig:PDFEL} (left panel), where the diagram of saturated intensity is depicted as a function of both parameters: It reveals that for $b_0$ below some $b_c\approx 0.3$, as well as for $\delta$ below $\delta_c\approx 2.3$ (top panels), the transition is sharp, whereas beyond these values, it turns out to be smooth. Thus, $(b_c,\delta_c)$ represents the critical values of parameters beyond which the transition from lasing to non-lasing turns into a smooth one: Although it is in no way supported by an entropic approach as it was for HMF, it can be seen as a dynamical version of the tricritical point present in the latter toy-model.

Regarding the possibilities to observe this peculiar phenomenology on the FERMI@Elettra FEL, a similar phase diagram has been plotted on Fig.\ref{fig:PDFEL} (middle panel), which accounts for the finite interaction length of the machine. It reveals that, despite the deep saturation may not be well established within the undulator length available, the two areas where a sharp vs. smooth transitions could be observed are quite well separated.

\section{The Collective Atomic Recoil Laser\label{sec:CARL}}

CARL consists of a collection of cold two-level atoms driven by a
far-detuned laser pump of frequency $\omega_p$ which radiates at
the frequency $\omega\sim\omega_p$ in the direction opposite to
the pump \cite{CARL}. In both the FEL and CARL systems the
radiation process arises from a collective instability which
originates from a symmetry breaking in the spatial distribution,
\textit{i.e.} a self-bunching of particles which group in regions
smaller than the wavelength. In the limit in which the radiation
pressure due to the pump laser can be neglected (for instance by
largely detuning the pump frequency from the atomic resonance),
the CARL is described  by the same dimensionless FEL equations
(\ref{eq:dynFEL})
\begin{eqnarray}
\frac{d\theta_j}{d\bar t}&=&p_j \label{C1}\\
\frac{d p_j}{d\bar t}&=&-\left(Ae^{i\theta_j}+A^*e^{-i\theta_j}\right)\label{C2}\\
\frac{dA}{d\bar t}&=&\frac{1}{N}\sum_{i=1}^N
e^{-i\theta_j}+i\delta A-\kappa A \label{C3},
\end{eqnarray}
however with the presence of a damping term $-\kappa A$ in the
field equation, accounting for radiation losses from a ring cavity
surrounding the atoms. Although CARL and FELs evolve with a
similar dynamics, the dimensionless variables of the two systems,
and consequently also the typical timescales, are very different.
In CARL, the phase and the normalized momentum of the atoms $j$
are $\theta_j=2k(z_j(t)-\langle v_z\rangle_0 t)$ and
$p_j=m(v_{zj}(t)-\langle v_z\rangle_0)/(2\hbar k\rho_C)$ (where
$z_j(t)$ and $v_{zj}(t)$ are the position and velocity of the $j$th
atom along the direction of the scattered field and $\langle
v_z\rangle_0$ is the average initial velocity), whereas $A$ stands
for the normalized complex amplitude of the radiation field,
$A=(\epsilon_0/\hbar\omega n_a\rho_C)^{1/2}E_0$, where $n_a$ is
the atomic density. The scaled time is $\bar
t=(8\omega_{rec}\rho_C)t$, where $\omega_{rec}=\hbar k^2/2m$ is
the recoil frequency, $\delta=(\omega-\omega_p-2k\langle
v_z\rangle_0)/(8\omega_{rec}\rho_C)$ is the pump-probe detuning,
$\kappa=\kappa_c/(8\omega_{rec}\rho_C)$ is the scaled loss of a
ring cavity with length $L_{cav}$, transmission $T$ and
$\kappa_c=cT/L_{cav}$, and finally $\rho_C=(\Gamma/8)(c\sigma_0
n_a/\Delta^2\omega_{rec}^2)^{1/3}(I/I_{sat})^{1/3}$, where
$\Gamma$ is the natural decay rate if the excited state,
$\sigma_0=3\lambda^2/2\pi$ is the scattering cross section,
$\Delta=\omega_0-\omega_p$ is the pump-atom detuning, $I$ is the
pump intensity and $I_{sat}=\hbar\omega\Gamma/2\sigma_0^2$ is the
saturation intensity.

There have been different experiments that have observed the CARL effect in room temperature gases \cite{bigelow}, cold atomic samples from Magneto-Optical traps (MOT) \cite{Tubing1, Tubing2} or Bose-Einstein condensates (BEC) \cite{inouye, rudi, Tubing3}.  CARL experiments were performed either in high-finesse optical cavities \cite{Tubing1} or in free space \cite{rudi}, where the effect was originally interpreted as Superradiant Rayleigh Scattering \cite{inouye}. However, it has later been emphasized that these experiments in free space can be seen as a CARL process in the superradiant regime \cite{CARL:cav,slama}.

\begin{figure}
\begin{center}
\includegraphics[width=9cm]{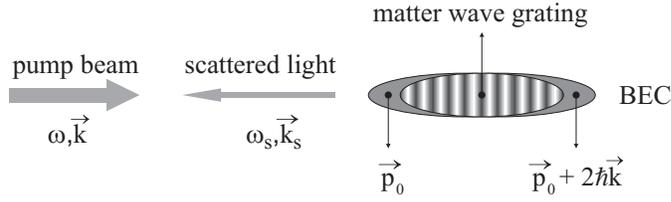}
\end{center}
\caption{Schematics of superradiant light scattering from a
Bose-Einstein condensate. An elongated BEC is illuminated by a far
off-resonant laser beam (pump beam) with frequency $\omega$ and
wavevector $\vec k$ directed along its axial direction. After
backscattering of photons with $\vec k_{s} \simeq -\vec k$ and the
subsequent recoil of atoms, a matter wave grating forms, due to the
quantum interference between the two momentum components of the
wavefunction of the condensate. The effect of this grating is to
further scatter the incident light in a self-amplifying process.}
\label{schemaCARL}
\end{figure}

At LENS, the CARL experiment~\cite{fallani} is realized with a cigar-shaped BEC of $^{87}$Rb
produced in a Ioffe-Pritchard magnetic trap by means of RF-induced
evaporative cooling. After 2 ms of free expansion, when the magnetic trap field is
completely switched off and the atomic cloud still has an elongated
shape (at this time the radial and axial sizes of the condensate are
typically 10 and 70 $\mu$m, respectively), a square pulse of light
is applied along the \textit{z}-axis (see Fig.\ref{schemaCARL}). The size of the laser
beams is larger than $0.5$ mm, far larger than the condensate free
fall during the interaction with light. In this geometry the CARL
process causes the pump light to be backscattered and the
self-amplified matter-wave propagates in the same direction as the
incident light.

This experiment allows for a great flexibility in the preparation of the initial state of the system: for example, it is possible to prepare the atoms in an initially bunched state by imposing an electromagnetic standing wave before the pump laser is activated. Regarding the momenta spread, it can be varied by cooling only partially the atoms. Finally, a non-zero pump-probe detuning $\delta$ can be induced by giving the atoms an initial momentum, which can be up to $1000 \hbar k$: Indeed to change the initial detuning of the probe, one can either change the initial atomic momentum or change the cavity frequency since conservation of energy imposes the frequency of the backscattered photon \cite{bonifacio04}.

\begin{table}[!ht]
\begin{center}
\caption{\label{tab:lensprms}Main parameters of LENS CARL.}
\begin{tabular}{@{}cccccccc}
\br
$\omega_{rec}$ & $\rho_C$ & t & $\delta$ & $\kappa$ & $\Delta p$ \\
\mr
5000Hz & 100-1000 & 1$\mu$s-5ms & -5 - 5 & 0.025-10$^5$ & 0.01-5 \\
\br
\end{tabular}
\end{center}
\end{table}

Equilibrium dynamical phase transitions were studied in the CARL context under different hypotheses: For example, using a thermalization hypothesis for the atoms, the CARL will reach equilibria where (equilibria) phase transitions can occur~\cite{jalavoyes03}. In the case where dissipation is counterbalanced by a stochastic Langevin force \cite{Tubing2}, steady states can be identified, as well as a phase transition analogous to the one that occurs in the Kuramoto model \cite{kuramoto}, yet with a self-generated collective oscillation frequency for the CARL. Both models were also connected when the strong friction is balanced by diffusion \cite{robb}, a regime for which steady states and phase transitions were also predicted \cite{Jalavoyes08}. Yet, the equilibrium condition remains a condition which is hard to satisfy experimentally. In this context, the study of the Quasi-Stationary regime, which are not $N$-body equilibrium regimes but rather stationary states of the associated Vlasov dynamics, is all the more relevant experimentally. 

Let's first consider the configuration where the CARL amplification is realized inside a ring cavity: the pump and probe light fields are counterpropagating modes of the ring cavity and the interaction time of the light fields with the atoms can be enhanced by several orders of magnitude, which supports the
amplification. Consequently, most of the CARL experiments carried out up to date employed ring cavities \cite{Tubing1,Tubing2,Tubing3}. Furthermore, in new experiments, especially in the microtrap-based ones, it appeared to be very profitable to introduce a high finesse optical cavity inside the vacuum system: Indeed, in this environment, Fabry-Perot cavities have already been demonstrated to yield very high finesses~\cite{brennecke}.

Back to the dynamical equations (\ref{C1} - \ref{C3}), it is clear that if the dissipation term $\kappa$ is small ($\kappa \ll 1$), the CARL is governed by the same equations as the FEL. In particular, when
monitoring the initial bunching and the detuning parameter, it should exhibit a dynamical transition similar to the one described in Sec.\ref{sec:FEL}. Simulations realized for a good cavity ($\kappa
=0.025$) reveal that the sharp/smooth transitions could indeed be observed in the relevant range of parameters (see Fig.\ref{fig:PDFEL}).

\begin{figure}[!ht]
\centerline{
$\begin{array}{cc}
\epsfig{figure=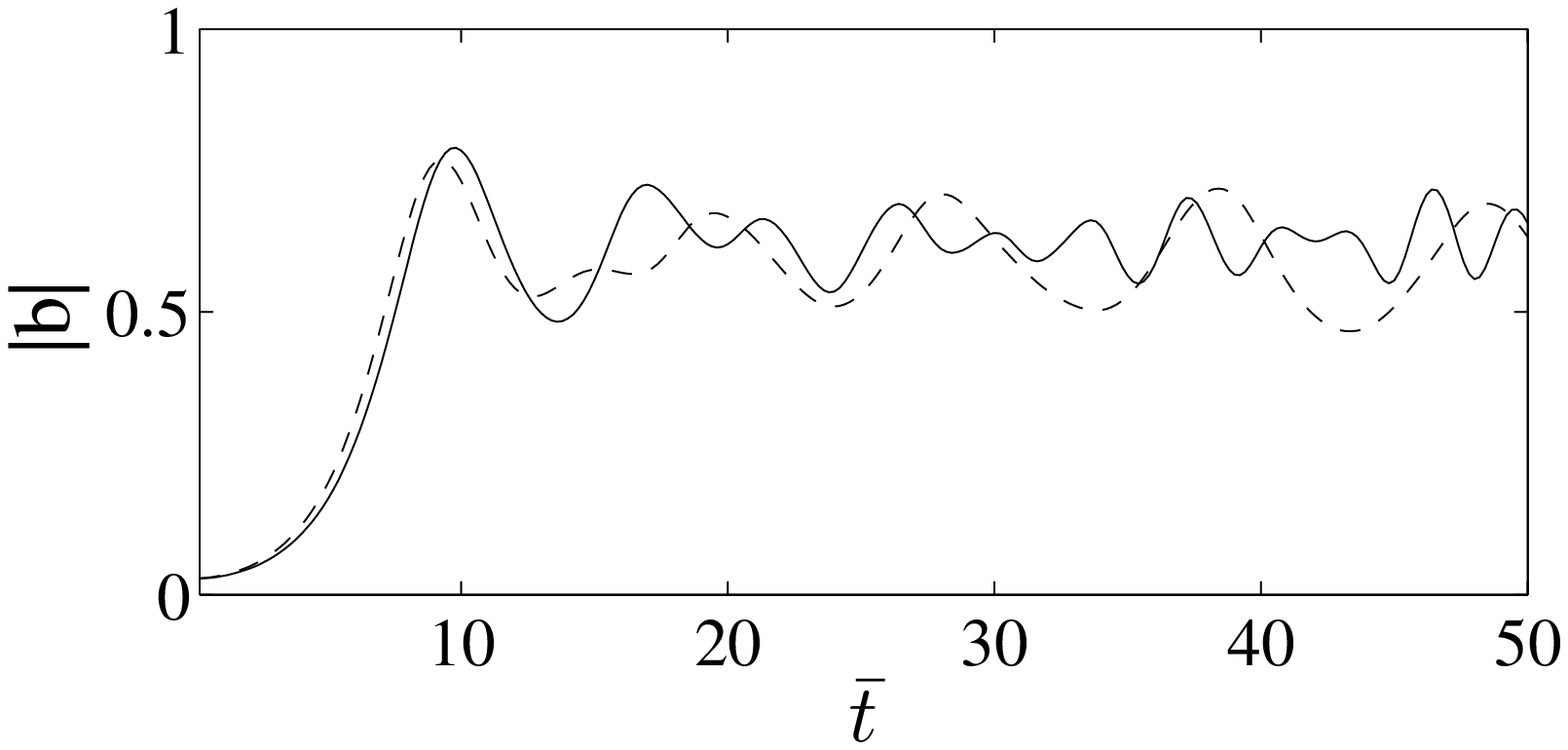,width=8cm}&\epsfig{figure=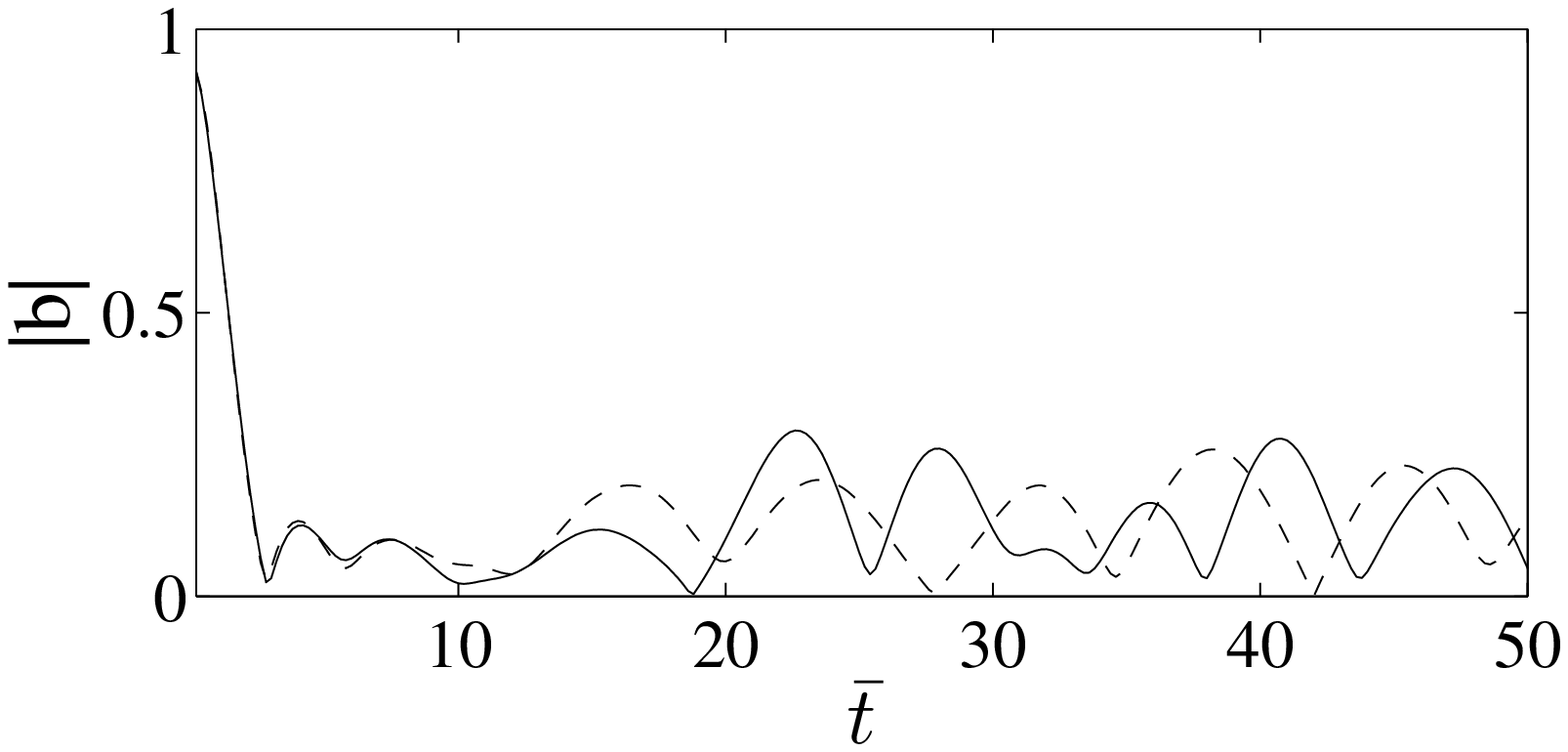,width=8cm}
\end{array}$}
\caption{Comparison between the dynamics of the CARL (plain lines) and HMF (dashed lines) dynamics, in magnetized regime (left: $b_0=0$, $\Delta \tilde{p}=0.1$) and unmagnetized regime (right: $b_0=0.94$, $\Delta \tilde{p}=1.5$). Simulations performed with $N=10000$ particles; CARL parameters: $\kappa=1$ and $\delta=-4$.\label{fig:bt}}
\end{figure}

On the other hand, the CARL experiment can be realized either without cavity, or within a low-quality one: This is at the cost of a high damping of the wave, modelled by the $\kappa>0$ term in Eq.(\ref{C3}). In this
case, if the dissipation is large enough ($\kappa \sim 1$), an
adiabatic treatment of the wave dynamics can be
performed~\cite{CARL:cav,fallani}, which corresponds to setting
$dA/dt=0$, which yields 
\beq
A=\frac{1}{\kappa-i\delta}\frac{1}{N}\sum_j e^{-i\theta_j}.\label{eq:adiab}
\eeq
Plugging this expression into the particles equations of motion
(\ref{C2}), we get the following equation for the momentum of
particle $j$ (the equation for its position is unchanged): \beq
\dot{p_j}=\frac{2}{\delta^2+\kappa^2}\frac{1}{N}\sum_m
\left(\delta\sin{(\theta_j-\theta_m)}-
\kappa\cos{(\theta_j-\theta_m)}\right). \label{eq:pjadiab} \eeq If
we now consider the extra limit where the detuning $\delta$ is
large with respect to $\kappa$ ($|\delta|\gg\kappa$),
Eq.(\ref{eq:pjadiab}) simply turns into \beq
\dot{p_j}=\frac{2}{\delta}\frac{1}{N}\sum_m
\sin{(\theta_j-\theta_m)}. \eeq Then, using the following
normalization
\begin{equation}
\begin{array}{rclrcl}
\tilde{\theta}_j&=&\theta_j, & \tilde{p}_j&=& p_j \sqrt{|\delta|/2}, \nonumber
\\ \tilde{t}&=&\bar t \sqrt{|\delta|/2}, & \tilde{H}&=& |\delta| H/2,\label{eq:norm}
\end{array}
\end{equation}
the system can be mapped into the HMF model (\ref{eq:hmf}). Anew,
$\epsilon=-sign(\delta)$ positive (negative) describes an
(anti)ferromagnetic interaction. Numerical simulations confirm
that in the above mentioned limit, despite small differences in
the dynamics, the CARL and HMF yield similar QSS regimes (see
Fig.\ref{fig:bt}).

\begin{figure}[!ht]
\centerline{
$\begin{array}{ccc}
\epsfig{figure=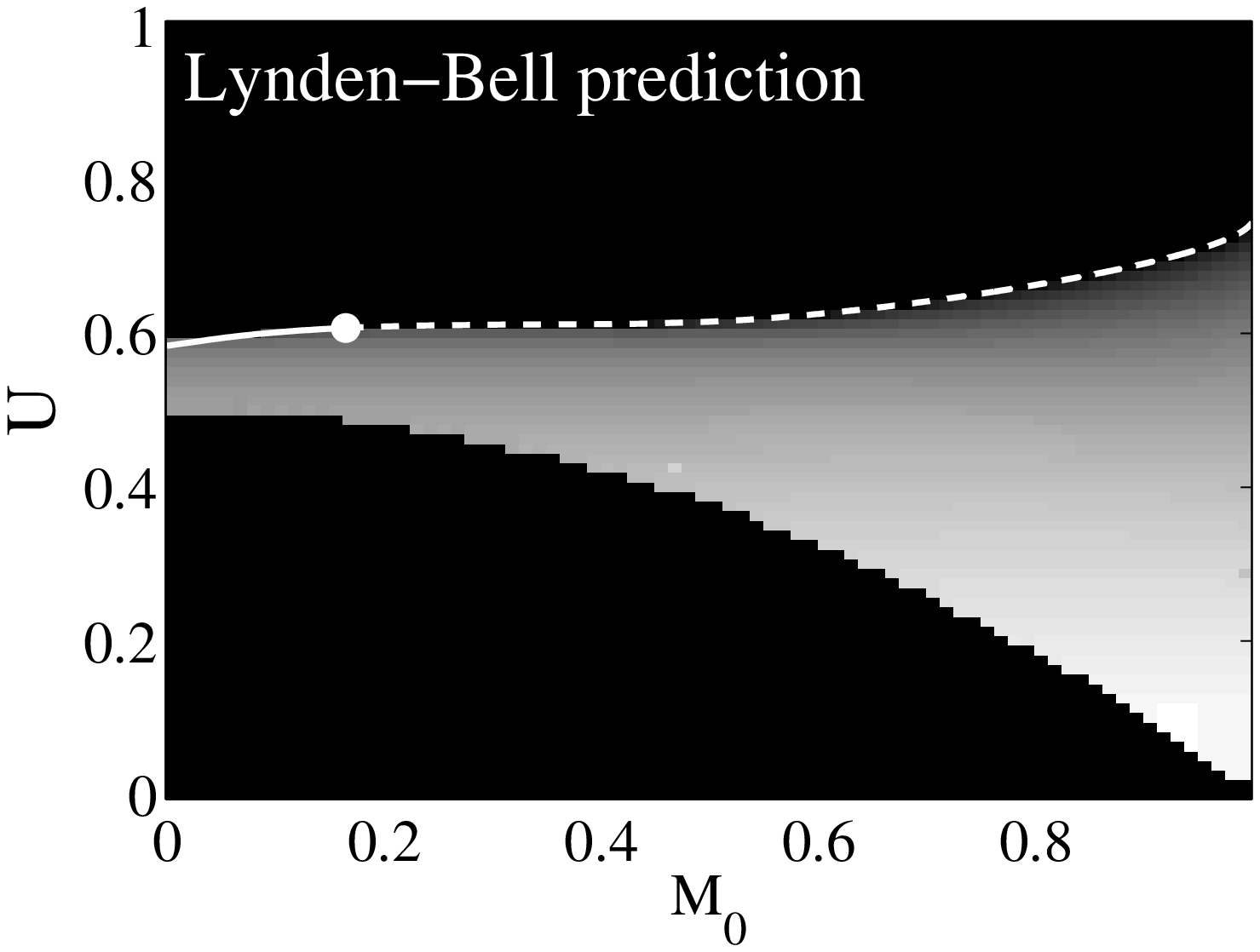,width=5cm}&\epsfig{figure=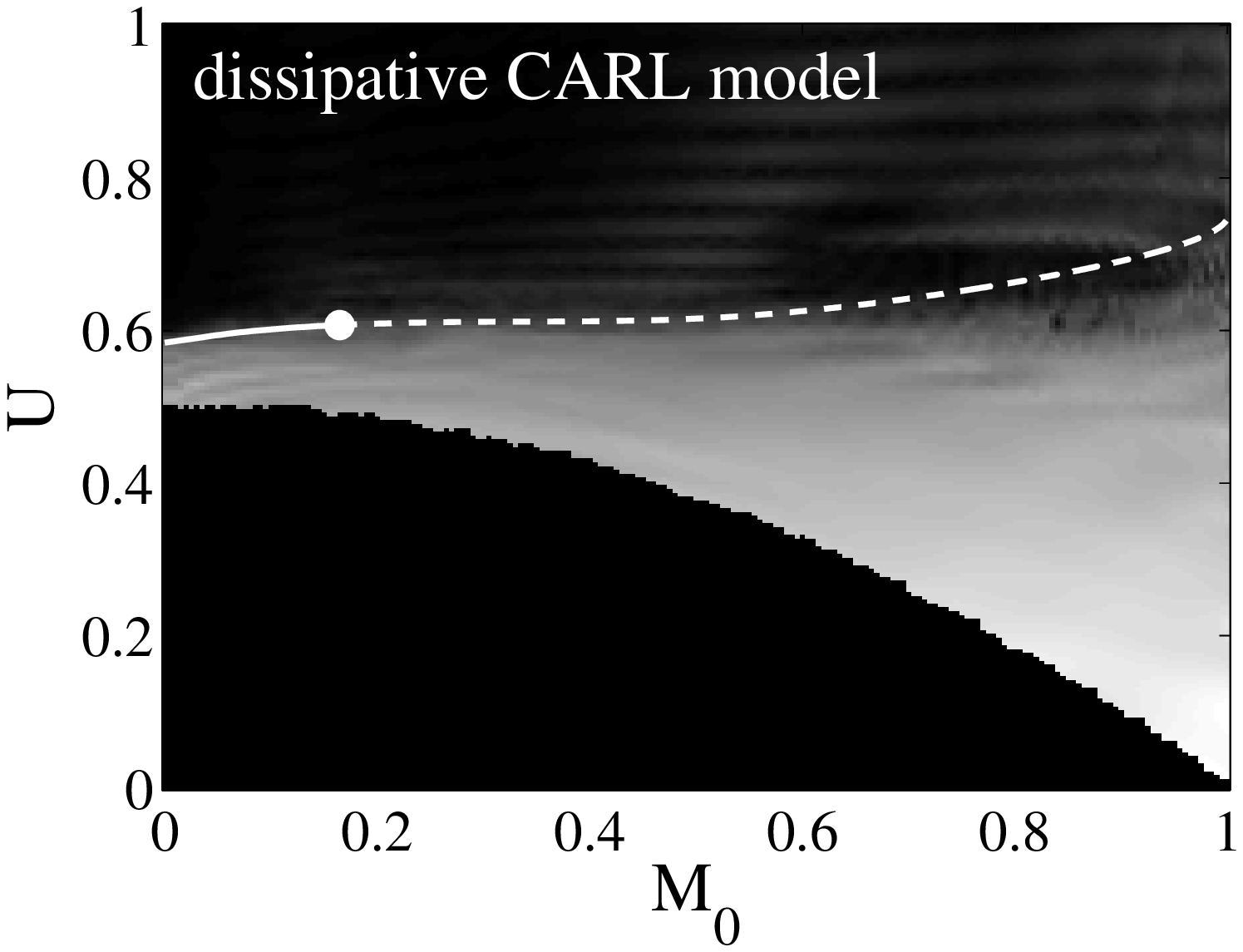,width=5cm}&\epsfig{figure=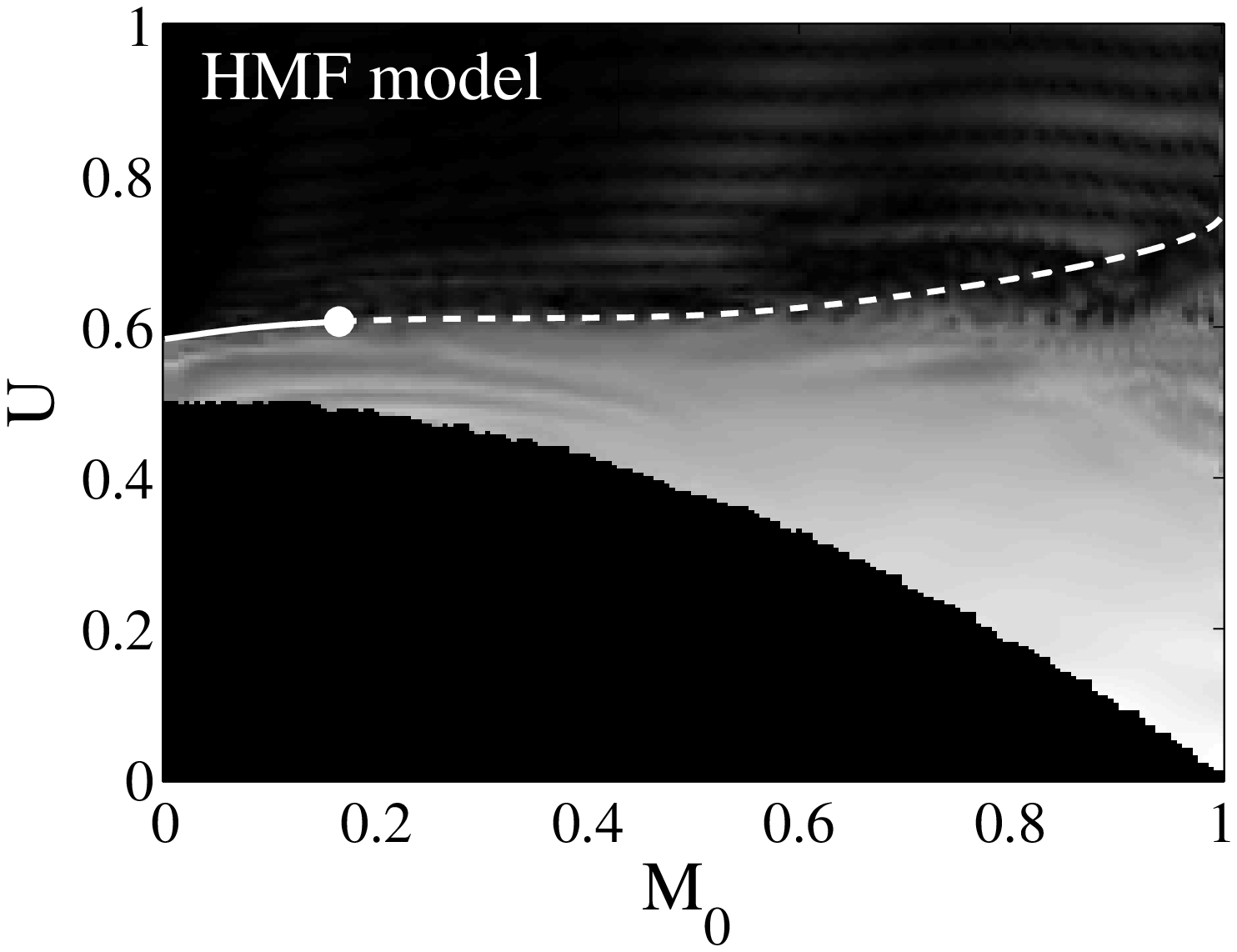,width=5cm}
\end{array}$}
\caption{Bunching factor as a function of the initial bunching and energy of the system, as predicted by Lynden-Bell's violent relaxation approach (left), and by $N$-body simulations of the CARL (center) and HMF (right) dynamics at finite length. The white line stands for the transition as predicted by Lynden-Bell, of the first order type below $M_c\approx 0.17$, and of the second order kind above. $N$-body simulations realized with $N=10000$ until $\bar{t}=40$; CARL simulations performed with $\kappa=0.5$ and $\delta=-5$. Note that the energy $U$ here refers to the {\it normalized} energy $U=\Delta\tilde{p}^2/6+(1-b_0^2)/2$.\label{fig:PD}}
\end{figure}

This formally links the CARL dynamics to the HMF model, in either its ferromagnetic or antiferromagnetic form, making in particular the bridge between the rich phenomenology predicted for the HMF model and a possible experimental realization. For example, regarding the ferromagnetic case, Fig.\ref{fig:PD} depicts the phase diagram of HMF in the $(b_0,U)$ plane as predicted by the LB prescription, and those obtained by direct $N$-body simulations of both the CARL dynamics and the HMF model. They reveal an excellent agreement between the CARL phase diagram and its reduced counterpart, the HMF, as well as with the Lynden-Bell approach. Note that the presence of fringes in the large $U$ part in the diagrams are due to the short time considered: yet, although the QSS regime is not well established, the transition line is already present and in good agreement with the LB prescription. This allows to conclude on the presence of an out-of-equilibrium phase transition in the CARL device, such as predicted for the HMF model~\cite{antoniazzi07b}.

\section{On the possibility to observe the QSS signature\label{sec:ccl}}

In this paper, we investigated the out-of-equilibrium dynamics of two long-range models which admit corresponding experimental implementations. We shall now
conclude the discussion by elaborating on the experimental possibilities to verify the correctness of the proposed picture and in particular detect the 
predicted phase transitions.

As concerns the FEL, the particles phase-space is not experimentally accessible since the electrons have relativistic energies. However, by monitoring the laser 
intensity it should be in principle possible to detect the transition, as demonstrated in Sec.\ref{sec:FEL}. Depending on the specific initial condition, ultimately characterized by an assigned electron bunching amount, the laser can gain in potency or, alternatively, keep its off mode. Experimentwise, by setting the initial bunching at either zero (no interaction in the modulator) or high values ($b_0\approx 0.4$), and tuning the electrons energy, one should in principle observe either a sharp or a smooth transition, as indicated by the simulations results. This is an interesting feature, indirect signature of the QSS existence, which bears an intringuing similarity
with the HMF behaviour. In this latter case, however, the presence of an out-of-equilibrium phase transition of both first and second order types was demonstrated on solid theoretical grounds. A similar theoretical justification is still lacking with reference to the FEL model. 

The CARL device,  in the regime of small dissipation, is predicted to exhibit a transition, similar to that displayed by the FEL. Again, and in analogy with the above, we   suggest that the presence of the transition could be successfully evidenced by recording the radiated intensity, under different experimental conditions. In addition, the CARL could also allow to access direct information on the particles dynamics in phase space: First, the atoms density can be recorded, and the presence of fringes used to quantify the degree of bunching. Then, letting the atoms expand after the CARL process ends, and recording their late positions, one can resolve the momentum distribution: Such a diagnostic would translate into an independent tool to bring into evidence the transition between distinct dynamical regimes. 

Finally, it is when operating the CARL device in the dissipative HMF-like regime that the possibility to measure the 
phase-space structures could be of paramount importance. The fringes, which means bunching, are in principle strongly correlated to the radiated power (see Eq.\ref{eq:adiab}). By accessing the distribution of momenta one could eventually detect the two bumps that are predicted to occur in presence of 
unbunched QSS, and which are believed to correspond to counter-propagating clusters of particles. More generally, and with reference to the  ferromagnetic case~\cite{antoniazzi07,barre09} we shall be interested in accurately investigating the overall velocity profile whose characteristics have been  
object of vigorous debates~\cite{latora,antoniazzi07c}.

\section{Acknowledgments}

The research of PdB is financially supported by the Belgian Federal Government (Interuniversity Attraction Pole ``Nonlinear systems, stochastic processes, and statistical mechanics'', 2007-2011). DF thanks financial support from the Euratom association. RB, GDN and FS aknowledge the scientific and financial support of the FERMI project.

\end{document}